%% file: main.tex
\documentclass[smallabstract,smallcaptions]{dccpaper}

\usepackage{amssymb}
\usepackage{amsthm}

\usepackage[mathletters]{ucs}
\usepackage[utf8x]{inputenc}

\usepackage{mathtools}

\usepackage{paralist}

\usepackage{booktabs}
\usepackage{tabu}
\newcolumntype{?}{!{\vrule width 2pt}}
\usepackage{multirow}
\usepackage{xcolor}
\usepackage{colortbl}

\usepackage{pgfplots}
\usepgfplotslibrary{groupplots}

\theoremstyle{plain}
\newtheorem{theorem}{Theorem}

\theoremstyle{definition}

\newtheorem{example}[theorem]{Example}
\theoremstyle{remark}

\ifx\numberwithinsect\relax
  \@addtoreset{theorem}{section}
  \expandafter\def\expandafter\thetheorem\expandafter{%
    \expandafter\thesection\expandafter\@thmcountersep\thetheorem}
\fi

\usepackage{refcount}

\usepackage{ascii} 

\usepackage{algorithm}
\usepackage[noend]{algpseudocode}
\algrenewcommand\alglinenumber[1]{\scriptsize #1:}
\usepackage{multicol}

\algnewcommand{\algorithmicand}{\textbf{and }}
\algnewcommand{\AND}{\algorithmicand}
\algnewcommand{\algorithmicor}{\textbf{or }}
\algnewcommand{\OR}{\algorithmicor}
\algnewcommand\algorithmiclet{\State \textbf{let }}
\algnewcommand\Let{\algorithmiclet}
\algnewcommand\algorithmicforeach{\textbf{for each}}
\algdef{S}[FOR]{ForEach}[1]{\algorithmicforeach\ #1\ \algorithmicdo}

\usepackage{tikz}
\usetikzlibrary{positioning,automata}
\usepackage{tikz-qtree}

\usetikzlibrary{arrows.meta,patterns}

\usepackage[labelfont=bf,font={small,it}]{caption}
\usepackage[os=mac, mackeys=symbols]{menukeys} 

\newcommand{\eox}{\hfill{\ensuremath{\Diamond}}}

\newcommand{\len}[1]{{\vert{#1}\vert}}
\newcommand{\pr}{\mathcal{R}}
\newcommand{\tuple}[1]{{\left\langle #1 \right\rangle}}
\newcommand{\lang}[1]{{\mathcal{L}(#1)}}
\newcommand{\produces}{\Rightarrow}

\def\tool#1{{{\asciifamily #1}}}
\newcommand{\winner}[1]{\textcolor{blue}{\textbf{#1}}}
\newcommand{\second}[1]{\textcolor{black}{\textbf{#1}}}
\newcommand{\loser}[1]{\textcolor{red}{\textbf{#1}}}

\newcommand{\True}{\text{\emph{true}}}
\newcommand{\False}{\text{\emph{false}}}
\usepackage{stmaryrd}
\SetSymbolFont{stmry}{bold}{U}{stmry}{m}{n}

\newcommand*{\nl}{\return} 

\newcommand*{\varleft}[1]{{\text{{\asciifamily L}}_{#1}}}
\newcommand*{\varright}[1]{{\text{{\asciifamily R}}_{#1}}}
\newcommand*{\varnl}[1]{{\text{{\asciifamily N}}_{#1}}}
\newcommand*{\varcount}[1]{{\text{{\asciifamily M}}_{#1}}}
\newcommand*{\counting}[1]{{\mathcal{C}_{#1}}}
\newcommand*{\edges}[1]{{\text{{\asciifamily E}}_{#1}}}

\newcommand*{\algvarleft}[1]{{\text{{\asciifamily L}}_{#1}}}
\newcommand*{\algvarright}[1]{{\text{{\asciifamily R}}_{#1}}}
\newcommand*{\algvarnl}[1]{{\text{{\asciifamily N}}_{#1}}}
\newcommand*{\algvarcount}[1]{{\text{{\asciifamily M}}_{#1}}}

\usepackage{tikz}
\usetikzlibrary{ipe} 

\tikzstyle{ipe stylesheet} = [
  ipe import,
  even odd rule,
  line join=round,
  line cap=butt,
  ipe pen normal/.style={line width=0.4},
  ipe pen heavier/.style={line width=0.8},
  ipe pen fat/.style={line width=1.2},
  ipe pen ultrafat/.style={line width=2},
  ipe pen normal,
  ipe mark normal/.style={ipe mark scale=3},
  ipe mark large/.style={ipe mark scale=5},
  ipe mark small/.style={ipe mark scale=2},
  ipe mark tiny/.style={ipe mark scale=1.1},
  ipe mark normal,
  /pgf/arrow keys/.cd,
  ipe arrow normal/.style={scale=7},
  ipe arrow large/.style={scale=10},
  ipe arrow small/.style={scale=5},
  ipe arrow tiny/.style={scale=3},
  ipe arrow normal,
  /tikz/.cd,
  ipe arrows, 
  <->/.tip = ipe normal,
  ipe dash normal/.style={dash pattern=},
  ipe dash dashed/.style={dash pattern=on 4bp off 4bp},
  ipe dash dotted/.style={dash pattern=on 1bp off 3bp},
  ipe dash dash dotted/.style={dash pattern=on 4bp off 2bp on 1bp off 2bp},
  ipe dash dash dot dotted/.style={dash pattern=on 4bp off 2bp on 1bp off 2bp on 1bp off 2bp},
  ipe dash normal,
  ipe node/.append style={font=\normalsize},
  ipe stretch normal/.style={ipe node stretch=1},
  ipe stretch normal,
  ipe opacity 10/.style={opacity=0.1},
  ipe opacity 30/.style={opacity=0.3},
  ipe opacity 50/.style={opacity=0.5},
  ipe opacity 75/.style={opacity=0.75},
  ipe opacity opaque/.style={opacity=1},
  ipe opacity opaque,
]

\definecolor{red}{rgb}{1,0,0}
\definecolor{green}{rgb}{0,1,0}
\definecolor{blue}{rgb}{0,0,1}
\definecolor{yellow}{rgb}{1,1,0}
\definecolor{orange}{rgb}{1,0.647,0}
\definecolor{gold}{rgb}{1,0.843,0}
\definecolor{purple}{rgb}{0.627,0.125,0.941}
\definecolor{gray}{rgb}{0.745,0.745,0.745}
\definecolor{brown}{rgb}{0.647,0.165,0.165}
\definecolor{navy}{rgb}{0,0,0.502}
\definecolor{pink}{rgb}{1,0.753,0.796}
\definecolor{seagreen}{rgb}{0.18,0.545,0.341}
\definecolor{turquoise}{rgb}{0.251,0.878,0.816}
\definecolor{violet}{rgb}{0.933,0.51,0.933}
\definecolor{darkblue}{rgb}{0,0,0.545}
\definecolor{darkcyan}{rgb}{0,0.545,0.545}
\definecolor{darkgray}{rgb}{0.663,0.663,0.663}
\definecolor{darkgreen}{rgb}{0,0.392,0}
\definecolor{darkmagenta}{rgb}{0.545,0,0.545}
\definecolor{darkorange}{rgb}{1,0.549,0}
\definecolor{darkred}{rgb}{0.545,0,0}
\definecolor{lightblue}{rgb}{0.678,0.847,0.902}
\definecolor{lightcyan}{rgb}{0.878,1,1}
\definecolor{lightgray}{rgb}{0.827,0.827,0.827}
\definecolor{lightgreen}{rgb}{0.565,0.933,0.565}
\definecolor{lightyellow}{rgb}{1,1,0.878}
\definecolor{black}{rgb}{0,0,0}
\definecolor{white}{rgb}{1,1,1}

\newcommand*\pct{\scalebox{.8}{\%}}

\usepackage[bottom]{footmisc} 
\usepackage{hyperref}
\hypersetup{hidelinks} 

\algrenewcommand\algorithmicindent{0.9em}%

\newcommand{\algSize}{\fontsize{10}{11}\selectfont} 
\usepackage{enumitem}

\makeatletter
\renewcommand{\paragraph}{%
  \@startsection{paragraph}{4}%
  {\z@}{1.5ex \@plus 1ex \@minus .2ex}{-1em}%
  {\normalfont\normalsize\bfseries}%
}
\makeatother

\newlength{\figurewidth}
\newlength{\smallfigurewidth}

\newenvironment{nscenter}
 {\parskip=0pt\par\nopagebreak\centering}
 {\par\noindent\ignorespacesafterend}

\begin{document}

\setlength{\abovedisplayskip}{5pt}
\setlength{\belowdisplayskip}{5pt}

\title{\large\textbf{Regular Expression Search on Compressed Text}}

\author{%
Pierre Ganty$^{\ast}$ and Pedro Valero$^{\ast\dag}$\\[0.5em]
{\small\begin{minipage}{\linewidth}\begin{center}
\begin{tabular}{ccc}
$^{\ast}$IMDEA Software Institute & \hspace*{0.5in} & $^{\dag}$Universidad Polit\'ecnica de Madrid \\
\url{pierre.ganty@imdea.org} && \url{pedro.valero@imdea.org}\\
\end{tabular}
\end{center}\end{minipage}}
}

\maketitle
\thispagestyle{empty}

\begin{abstract}
We present an algorithm for searching regular expression matches in compressed text.
The algorithm reports the number of matching lines in the uncompressed text in time \emph{linear} in the size of its \emph{compressed} version.
We define efficient data structures that yield \emph{nearly optimal} complexity bounds and provide a \emph{sequential} implementation --\tool{zearch}-- that requires up to $25\pct$ less time than the state of the art.
\end{abstract}

\Section{Introduction}
The growing amount of information handled by modern systems demands efficient techniques both for compression, to reduce the storage cost, and for regular expression searching, to speed up querying. 
A type of query that is supported out of the box by many tools\footnote{Tools such as \tool{grep}, \tool{ripgrep}, \tool{awk} and \tool{ag}, among others, can be used to report the number of matching lines in a text.} is \emph{counting}: compute how many lines of the input text contain a match for the expression.   
When the text is given in compressed form, the fastest approach in practice is to query the uncompressed text as it is recovered by the decompressor.

We present an algorithm for counting the lines in a compressed text containing a match for a regular expression whose runtime does not depend on the size $N$ of the uncompressed text.
Instead, it runs in time \emph{linear} in the size of its \emph{compressed version}.
Furthermore, the information computed for counting can be used to perform an \emph{on-the-fly}, \emph{lazy} decompression to recover the matching lines from the compressed text.
Note that, for reporting the matching lines, the dependency on $N$ in unavoidable.

The salient features of our approach are:
\paragraph{Generality.} Our algorithm is not tied to any particular grammar-based compressor.
Instead, we consider the compressed text is given by a straight line program (SLP): a context-free grammar generating the uncompressed text and nothing else.

Finding the smallest SLP $g$ generating a text of length $N$ is an NP-hard problem, as shown by Charikar et al.~\cite{Charikar2005Smallest}, for which grammar-based compressors such as LZ78~\cite{ziv1978compression}, LZW~\cite{welch1984technique}, RePair~\cite{larsson2000off} and Sequitur~\cite{nevill1997compression} produce different approximations.
For instance, Hucke et al.~\cite{Hucke2016Smallest} showed that the LZ78 algorithm produces a representation of size $Ω\bigl(\len{g}{\cdot} (N/\log{N})^{2/3}\bigr)$ and the representation produced by the RePair algorithm has size $Ω\bigl(\len{g}{\cdot} (\log{N}/\log\log{N})\bigr)$.
Since it is defined over SLPs, our algorithm applies to all such approximations, including $g$ itself.

\paragraph{Nearly optimal data structures.} We define data structures enabling the algorithm to run in time linear in the size of the compressed text. \pagebreak
With these data structures our algorithm runs in $\mathcal{O}(p {\cdot} s^3)$ time using $\mathcal{O}(p {\cdot} s^2)$ space where $p$ is the size of the compressed text and $s$ is the size of the automaton built from the expression. 
When the automaton is deterministic, the complexity drops to $\mathcal{O}(p{\cdot}s)$ time and $\mathcal{O}(p{\cdot}s)$ space.
Abboud et al.~\cite{Amir2018FineGrained} showed that there is no combinatorial\footnote{Interpreted as
any \emph{practically efficient} algorithm that does not suffer from the issues of Fast Matrix Multiplication such as large constants and inefficient memory usage.} algorithm improving these\linebreak time complexity bounds beyond \emph{polylog} factors, hence our algorithm is \emph{nearly optimal}.

\paragraph{Efficient implementation.} We present \tool{zearch}, a purely \emph{sequential} implementation of our algorithm which uses the above mentioned data structures.\footnote{\tool{zearch} can optionally report the matching lines.}
The experiments show that \tool{zearch} requires up to $25\pct$ less time than the state of the art: running \tool{hyperscan} on the uncompressed text as it is recovered by \tool{lz4} (in \emph{parallel}).
Furthermore, when the grammar-based compressor achieves high compression ratio (above 13:1), running \tool{zearch} on the compressed text is as fast as running \tool{hyperscan} directly on the uncompressed text.
This is the case, for instance, when working with automatically generated log files.

\Section{Notation}\label{sec:preliminaries}
An \emph{alphabet} \(Σ\) is a nonempty finite set of \emph{symbols}.
A \emph{string} \(w\) is a finite sequence of symbols of \(Σ\) where the empty sequence is denoted \(ε\).
Let \(\len{w}\) denote the \emph{length} of \(w\) that we abbreviate to \(†\) when \(w\) is clear from the context.
Further define \( (w)_i \) as the \(i\)-th symbol of \(w\) if \(1 ≤ i ≤ †\) and \(ε\) otherwise.
Similarly, \((w)_{i,j}\) denotes the substring, also called \emph{factor}, of $w$ between the $i$-th and the $j$-th symbols, both included.


A \emph{finite state automaton} (FSA or automaton for short) is a tuple $A=(Q,Σ,I,F,δ)$ where $Q$ is the (finite) set of \emph{states}; $I \subseteq Q$ are the \emph{initial states}; $Σ$ is the alphabet; $F \subseteq Q$ are the \emph{final states}; and $δ \subseteq Q \times Σ \times Q$ are the \emph{transitions}\footnote{Our definition prescribes $ε$-transitions which can be removed in $\mathcal{O}(s^2)$ time adding no state.}.
The notions of \emph{accepting run}, \emph{accepted string} and \emph{language} of an automaton (denoted \(\lang{A}\)) are defined as expected. 
For clarity, we assume through the paper that $I$ and $F$ are disjoint.
Otherwise $ε \in \lang{A}$, therefore, for every string there is a factor ($ε$) in $\lang{A}$.


A \emph{Straight Line Program}, hereafter SLP, is a tuple $P=(V,Σ,\pr)$ where $V$ is the set of \emph{variables} $\{X_1,X_2,…,X_{\len{V}}\}$, $Σ$ is the alphabet and $\pr$ is the set of rules $\{X_i → α_i β_i \mid α_i,β_i \in (Σ \cup \{X_1,…,X_{i-1}\}\}$.
We refer to $X_{\len{V}} → \alpha_\len{V}β_\len{V}$ as the \emph{axiom rule}.
We write \(ρ ⇒ σ\), with $ρ,σ \in (Σ \cup V)^*$, if there exists \( (ρ)_i = X\) and \((X → αβ)\in \pr\) such that \( σ = (ρ)_{1,i-1}αβ (ρ)_{i+1,\dag}\).
Denote the reflexo-transitive closure of \(\Rightarrow\) by $⇒^*$.
Clearly, the language generated by an SLP consists of a single string $w ∈ Σ^*$ and, by definition, $\len{w} > 1$.
Since \(\lang{P}=\{w\}\) we identify \(w\) with \(\lang{P}\).


\Section{Counting Algorithm}\label{sec:provingCorrectness}
Let $\nl$ denote the new-line delimiter and $\widehat{Σ} = Σ {\setminus} \{\nl\}$.
Given a string $w \in Σ^+$ compressed as an SLP $P=(V,Σ,\pr)$ and an automaton $A=(Q,\widehat{Σ},I,F,δ)$ built from a regular expression, Algorithm~\ref{alg:algorithmTheoryCount} reports the number of lines in $w$ containing a match for the expression.
Note that we deliberately ignore matches across lines.

As an overview, our algorithm computes some \emph{counting information} for each alphabet symbol of the grammar (procedure \textsc{init\_automaton}) which is then propagated, in a bottom-up manner, to the axiom rule.
Such propagation is achieved by iterating through the grammar rules (loop in line~\ref{alg:algorithmTheoryCount:step}) and combining, for each rule, the information for the symbols on the right hand side to obtain the information for the variable on the left (procedure \textsc{count}).
Finally, the output of the counting is computed from the information propagated to the axiom symbol (line~\ref{alg:algorithmTheoryCount:return}).

\begin{algorithm}[!ht]
\algSize
\caption{Counting matching lines.}\label{alg:algorithmTheoryCount}
\textbf{Input:} An SLP $P=(V,Σ,\pr)$ and an FSA $A=(Q,\widehat{Σ},I,F,δ)$.

\textbf{Output:} The number of matching lines in $\lang{P}$.
\vspace{-8pt}
\begin{multicols}{2}
\begin{algorithmic}[1]
\Procedure{count}{$X$, $α$, $β$, $m$}
\State $\algvarnl{X}{:=}\algvarnl{α} \lor \algvarnl{β}$;

\State \(\algvarleft{X} {:=} ( \neg \algvarnl{α} \mathbin{?} \algvarleft{α} \lor \algvarleft{β} \lor m \colon \algvarleft{α} )\);
\State \(\algvarright{X} {:=} ( \neg \algvarnl{β} \mathbin{?} \algvarright{α}\lor \algvarright{β}\lor m \colon \algvarright{β} )\);
\State \(\algvarcount{X} {:=} \algvarcount{α}{+}\algvarcount{β}  {+} \bigl( \algvarnl{α} {\land} \algvarnl{β} {\land} (\algvarright{α} {\lor} \algvarleft{β} {\lor} m) \;{\mathbin{?}}\; {1}{\colon}{0}\bigr)\);
\EndProcedure

\Procedure{init\_automaton}{$\;$}
  \ForEach{$a \in Σ$}
    \State \(\algvarnl{a} {:=} (a = \nl) \);
    \(\algvarcount{a} {:=} 0 \);
    \State \(\algvarleft{a} {:=} \bigl((q_0,a,q_f) \in δ, \; q_0 \in I,\; q_f \in F\bigr)\);
    \State \(\algvarright{a} {:=} \algvarleft{a}\);
  \EndFor
\EndProcedure
\end{algorithmic}
\end{multicols}
\vspace{-13pt}
\begin{algorithmic}[1]
\setcounter{ALG@line}{10}
\Function{main}{}
  \State \Call{init\_automaton}{$\;$}
  \ForEach{$\ell = 1,2,\ldots,\len{V}{-}1$}\label{alg:algorithmTheoryCount:step}
    \Let $(X_\ell → α_\ell β_\ell) \in \pr$; \label{alg:algorithmTheoryCount:loopl}
    new\_match {:=} $\False$;
    \ForEach{$q_1,q'\in Q$ s.t. $(q_1,α_\ell,q')\in δ$ or $q_1{=}q'\in I$}\label{alg:algorithmTheoryCount:loopa}
      \ForEach{$q_2 \in Q$ s.t. $(q',β_\ell,q_2)∈ δ$ or $q'{=}q_2 \in F$} \hspace{10pt}\label{alg:algorithmTheoryCount:loopb}
        \State $δ{:=} δ \cup \{(q_1,X_\ell,q_2)\}$; \label{alg:algorithmTheoryCount:add}
        \State new\_match {:=} new\_match $\lor \bigl( q_1\in I\land q'\notin \bigl(I\cup F\bigr) \land q_2\in F\bigr)$\label{alg:algorithmTheoryCount:nm};
      \EndFor
    \EndFor
    \State \Call{count}{$X_\ell,α_\ell,β_\ell,$new\_match};
  \EndFor\label{alg:algorithmTheoryCount:looplend}
  \State \Return \(\algvarcount{X}_{\len{V}} + (\algvarnl{X}_{\len{V}} \ \mathbin{?}\ \algvarleft{X}_{\len{V}}{+}\algvarright{X}_{\len{V}} \colon \algvarleft{X}_{\len{V}})\);\label{alg:algorithmTheoryCount:return}
\EndFunction
\end{algorithmic}
\end{algorithm}
\vspace{-5pt}

Define a \emph{line} as a maximal factor of $w$ each symbol of which belongs to $\widehat{Σ}$, a \emph{closed line} as a line which is not a prefix nor a suffix of $w$ and a \emph{matching line} as a line in $\widehat{\lang{A}}$, where $\widehat{\lang{A}} = \widehat{Σ}^*{\cdot}\lang{A}{\cdot}\widehat{Σ}^*$.
The \emph{counting information of} $τ \in (V \cup Σ)$, with $τ\produces^* u$ and $u \in Σ^+$, is the tuple $\counting{τ}=\tuple{\varnl{τ},\varleft{τ},\varright{τ},\varcount{τ}}$ where
\begin{align*}
\varnl{τ} &:= \exists k\; (u)_k = \nl & 
\varleft{τ} & := \exists i \; (u)_{1,i} \in \widehat{Σ}^*{\cdot}\lang{A} \\
\varright{τ} &:= \exists j \; (u)_{j,\dag} \in \lang{A}{\cdot}\widehat{Σ}^*& 
\varcount{τ} & := \len{\{(i{+}1,j{-}1) \mid (u)_{i,j} \in \nl{\cdot}\widehat{\lang{A}}{\cdot}\nl\}}
\end{align*}

Note that $\varnl{τ}$, $\varleft{τ}$ and $\varright{τ}$ are boolean values while $\varcount{τ}$ is an integer.
It follows from the definition that the number of \emph{matching lines} in $u$, with $τ \produces^*u$, is given by the number of \emph{closed matching lines} ($\varcount{τ}$) plus the prefix of $u$ if{}f it is a \emph{matching line} ($\varleft{τ}$) and the suffix of $u$ if{}f it is a \emph{matching line} ($\varright{τ}$) different from the prefix ($\varnl{τ})$.
Since whenever $\varnl{τ} = \False$ we have $\varleft{τ} = \varright{τ}$, it follows that
\[\sharp\text{\emph{matching lines} in } u = \varcount{τ} + \left\{ 
\begin{array}{ll}
  1 & \text{if } \varleft{τ}\\
  0 & otherwise \end{array} \\
   \right. + \left\{ 
\begin{array}{ll}
  1 & \text{if } \varnl{τ} \land \varright{τ}\\
  0 & otherwise \end{array} \\
   \right.
\]

Computing the counting information of $τ$ requires deciding membership of certain factors of $u$ in $\widehat{\lang{A}}$.
To solve such queries we adapt an algorithm of Esparza et al.~\cite{esparza2000uniform} designed to decide whether the languages generated by a context-free grammar and an automaton intersect.
The resulting algorithm iterates through the rules $(X{→}αβ)\in\pr$ applying the following operation: add $(q_1,X,q_2)$ to $δ$ if{}f
\begin{inparaenum}[\upshape(\itshape a\upshape)]
\item $(q_1,α,q'),(q',β,q_2)\in δ$, 
\item $(q_1,β,q_2)\in δ \text{ with } q_1 \in I$, or
\item $(q_1,α,q_2) \in δ \text{ with } q_2 \in F$.
\end{inparaenum}
This operation corresponds to lines~\ref{alg:algorithmTheoryCount:loopa} to \ref{alg:algorithmTheoryCount:add} of Algorithm~\ref{alg:algorithmTheoryCount}.
As a result, after processing the rule for $τ$, we have $(q_1,τ,q_2) \in δ$ if{}f the automaton moves from $q$ to $q'$ reading \begin{inparaenum}[\upshape(\itshape a\upshape)]
\item $u$,
\item a suffix of $u$ and $q_1 \in I$, or
\item a prefix of $u$ and $q_2 \in F$.
\end{inparaenum}

Procedures \textsc{count} and \textsc{init\_automaton} are quite straightforward, the main difficulty being the computation of $\algvarcount{X}$ which we explain next.
Let $x,y \in Σ^+$ be the strings generated by $α$ and $β$, respectively.
Given rule $X{→}αβ$, $X$ generates all the matching lines generated by $α$ and $β$ plus, possibly, a ``new'' matching line of the form $z{=}(x)_{i,\dag}(y)_{1,j}$ with $1{<} i {\leq} \len{x}$ and $1 {\leq} j {<} \len{y}$.
Such an extra matching line appears if{}f both $α$ and $β$ generate a $\nl$ symbol and either the suffix of $x$ or the prefix of $y$ matches the expression or there is a new match $m \in z$ with $m \notin x$, $m \notin y$ (line~\ref{alg:algorithmTheoryCount:nm}).

\begin{example}
Let $A$ be an automaton with $\lang{A}=\{ab,ba\}$.
Consider the grammar rule $X{→}αβ$ with $α\produces^* ba\nl a$ and $β\produces^*b \nl aba$. 
Then $X\produces^* ba\nl ab \nl aba$.

The matching lines generated by $α$, $β$ and $X$ are, respectively, $\{ba\}$, $\{aba\}$ and $\{ba,ab,aba\}$.
Furthermore $\counting{α}{=}\tuple{\True,\True,\False,0}$ and $\counting{β}{=}\tuple{\True,\False,\True,0}$.

Applying function \textsc{count} we find that $\counting{X}=\tuple{\True,\True,\True,1}$.
Therefore the number of matching lines is $1{+}1{+}1{=}3$, as expected. \eox
\end{example}

Note that the counting information computed by Algorithm~\ref{alg:algorithmTheoryCount} can be used to uncompress \emph{only} the matching lines by performing a top-down processing of the SLP. 
For instance, given $X{→}αβ$ with $\counting{X}=\tuple{\True,\True,\False,0}$ and $\counting{α}=\tuple{\True,\True,\False,0}$, there is no need to decompress the string generated by $β$ since we are certain it is not part of any matching line (otherwise we should have $\varcount{X}>0$ or $\varright{X}=\True$).

Next, we describe the data structures used to implement Algorithm~\ref{alg:algorithmTheoryCount} with \emph{nearly optimal} complexity.

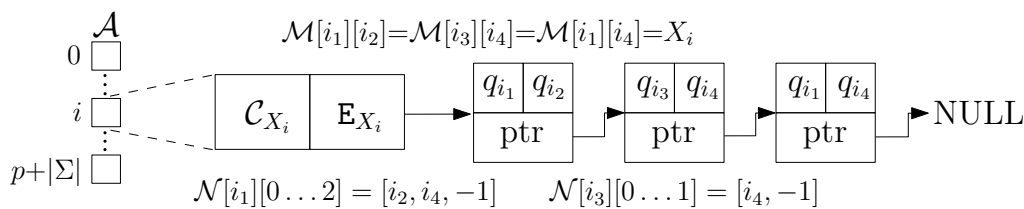
\begin{figure}[!ht]
\centering
\resizebox{0.9\textwidth}{!}{
\begin{tikzpicture}[ipe stylesheet]
\input{dataStructure.tex}
\end{tikzpicture}
}
\vspace{2pt}
\caption{Data structures enabling nearly optimal running time for Algorithm~\ref{alg:algorithmTheoryCount}. The image shows the contents of $\mathcal{M}$ after processing rule $X_i → α_i β_i$ and the contents of $\mathcal{N}$ after processing $X_\ell → α_\ellβ_\ell$ with $β_\ell = X_i$.}\label{fig:datastructure}
\vspace{-10pt}
\end{figure}

\SubSection{Data Structures}
We assume the alphabet symbols, variables and states are indexed and use the following data structures, illustrated in Figure~\ref{fig:datastructure}: an array $\mathcal{A}$ with $p{+}\len{Σ}$ elements, where $p$ is the number of rules of the SLP, and two $s \times s$ matrices $\mathcal{M}$ and $\mathcal{N}$ where $s$ is the number of states of the automaton.

\vspace{-7pt}

Each element $\mathcal{A}[i]$ contains the information related to variable $X_i$, i.e. $\counting{X_i}$ and the list of transitions labeled with $X_i$, $\edges{X_i}$.
We store $\counting{X}$ using one bit for each $\algvarnl{X}$, $\algvarleft{X}$ and $\algvarright{X}$ and an integer for $\algvarcount{X}$.
For each rule $X_\ell → α_\ell β_\ell$ the matrix $\mathcal{N}$ is set so that row $i$ contains all the states reachable from $q_i$ with a transition labeled with $β_\ell$, i.e. $\{q_j \mid (q_i,β_\ell,q_j) \in δ\}$.
If there are less than $s$ such states we use a sentinel value (${-}1$ in Figure~\ref{fig:datastructure}).
Finally, each element $\mathcal{M}[i][j]$ stores the index $\ell$ of the last variable for which $(q_i,X_\ell,q_j)$ was added to $δ$.
Note that since rules are processed one at a time, matrices $\mathcal{N}$ and $\mathcal{M}$ can be reused for all rules.

These data structures provide $\mathcal{O}(1)$ runtime for the following operations: 
\begin{compactitem}
\item Accessing the information corresponding to $α_\ell$ and $β_\ell$ at line~\ref{alg:algorithmTheoryCount:loopl} (using $\mathcal{A}$).
\item Accessing the list of pairs $(q,q')$ with $(q,α_\ell,q') \in δ$ at line~\ref{alg:algorithmTheoryCount:loopa} (using $\edges{X_i}$).
\item Accessing the list of states $q_2$ with $(q',β_\ell,q_2) \in δ$ at line~\ref{alg:algorithmTheoryCount:loopb} (using $\mathcal{N}$).
\item Inserting a pair $(q,q')$ in $\edges{X_i}$ at line~\ref{alg:algorithmTheoryCount:add} (using $\mathcal{M}$).
\end{compactitem}

As a result, Algorithm~\ref{alg:algorithmTheoryCount} runs in $\mathcal{O}(p{\cdot}s^3)$ time using $\mathcal{O}(p{\cdot}s^2)$ space when the FSA built from the regular expression is non deterministic and it runs in $\mathcal{O}(p{\cdot}s)$ time and $\mathcal{O}(p{\cdot}s)$ space when the FSA is deterministic (each row of $\mathcal{N}$ stores up to one state).

Abboud et al. proved~\cite[Thm.~3.2]{Amir2018FineGrained} that, under the Strong Exponential Time Hypothesis, there is no combinatorial algorithm deciding whether a grammar-compressed text contains a match for a deterministic FSA running in $\mathcal{O}((p {\cdot} s)^{1-ε})$ time  with $ε{>}0$.
For non deterministic FSA, they proved~\cite[Thm.~4.2]{Amir2018FineGrained} that, under the $k$-Clique Conjecture, there is no combinatorial algorithm running in $\mathcal{O}((p {\cdot} s^3)^{1-ε})$ time.
Therefore, our algorithm is \emph{nearly optimal} both for deterministic and non deterministic FSA.


\Section{Implementation}\label{sec:implementation}
We implemented Algorithm~\ref{alg:algorithmTheoryCount}, using the data structures described in the previous section, in a tool named \tool{zearch}\footnote{\url{https://github.com/pevalme/zearch}}.
This tool works on \tool{repair}\footnote{\url{https://storage.googleapis.com/google-code-archive-downloads/v2/code.google.com/re-pair/repair110811.tar.gz}}-compressed text and, beyond counting the matching lines, it can also report them by partially decompressing the input file.
The implementation consists of less than 2000 lines of C code.

The choice of this particular compressor, which implements the RePair algorithm~\cite{larsson2000off}, is due to the little effort required to adapt Algorithm~\ref{alg:algorithmTheoryCount} to the specific grammar built by \tool{repair} and the compression it achieves (see Table~\ref{table:compression}).
However \tool{zearch} can handle any grammar-based scheme by providing a way to recover the SLP from the input file.
Recall that we assume the alphabet symbols, variables and states are indexed. 
For \tool{repair}-compressed text, the indexes of the alphabet symbols are $0,1,…,255$ ($Σ$ is fixed\footnote{Our algorithm also applies to larger alphabets, such as UTF8, without altering its complexity}) and the indexes of the variables are $256…p{+}256$.
Grammar-based compressors encode the grammar so that rule $X{→}αβ$ appears always after the rules with $α$ and $β$ on the left hand side.
Thus, each iteration of the loop in line~\ref{alg:algorithmTheoryCount:loopl} reads a subsequent rule from the compressed input file.

We translate the input regular expression into an $ε$-free FSA using the automata library \tool{libfa}\footnote{\url{http://augeas.net/libfa/index.html}} which applies Thompson's algorithm~\cite{thompson1968programming} with on-the-fly $ε$-removal.


\Section{Empirical Evaluation}\label{sec:experimental}
Next we present a \emph{summary} of the experiments carried out to assess the performance of \tool{zearch}.
The details of the experiments, including the runtime and number of matching lines reported for each expression on each file and considering more tools, file sizes and regular expressions are available on-line\footnote{\url{https://pevalme.github.io/zearch/graphs/index.html}}.

All tools for regular expression searching considered in this benchmark are used to count the matching lines without reporting them.
To simplify the terminology, we refer to counting the matching lines as \emph{searching}, unless otherwise stated.

\SubSection{Tools}
Our benchmark compares the performance of \tool{zearch} against the fastest implementations we found for \begin{inparaenum}[\upshape(\itshape i\upshape)]
\item searching the compressed text without decompression,
\item searching the uncompressed text,
\item decompressing the text without searching and \linebreak
\item searching the uncompressed text as it is recovered by the decompressor.
\end{inparaenum}

For searching the compressed text we consider \tool{GNgrep}, the tool developed by Navarro~\cite{navarro2003regular} for searching \tool{LZW}-compressed text.
To the best of our knowledge, this is the only existing tool departing from the \emph{decompress and search} approach.

For searching uncompressed text we consider \tool{grep} and \tool{hyperscan}.
We improve the performance of \tool{grep} by compiling it without \emph{perl regular expression} compatibility, which is not supported by \tool{zearch}.
We used the library \tool{hyperscan} by means of the tool (provided with the library) \tool{simplegrep}, which we modified\footnote{\url{https://gist.github.com/pevalme/f94bedc9ff08373a0301b8c795063093}} to \emph{efficiently} read data either from stdin or an input file.

For (de)compressing the files we use \tool{zstd} and \tool{lz4} which are among the best lossless compressors\footnote{\url{https://quixdb.github.io/squash-benchmark/}}, being \tool{lz4} considerably faster while \tool{zstd} achieves better compression.
We use both tools with the highest compression level, which has little impact on the time required for decompression.

We use versions \tool{grep v3.3}, \tool{hyperscan v5.0.0}, \tool{lz4 v1.8.3} and \tool{zstd v1.3.6} running in an Intel Xeon E5640 CPU 2.67 GHz with 20 GB RAM which supports SIMD instructions up to SSE4-2.
We restrict to ASCII inputs and set \verb!LC_ALL=C! for all experiments, which significantly improves the performance of \tool{grep}.
Since both \tool{hyperscan} and \tool{GNgrep} count positions of the text where a match ends, we extend each regular expression (when used with these tools) to match the whole line.
We made this decision to ensure all tools solve the same counting problem and therefore produce the \emph{same output}.

\SubSection{Files and Regular Expressions}
Our benchmark consists of an automatically generated \emph{Log}\footnote{\url{http://ita.ee.lbl.gov/html/contrib/NASA-HTTP.html}} of HTTP requests, English \emph{Subtitles}~\cite{openSubtitles}, and a concatenation of English \emph{Books}\footnote{\url{https://web.eecs.umich.edu/~lahiri/gutenberg_dataset.html}}.
Table~\ref{table:compression} shows how each compressor behaves on these files.
\begin{table}[!ht]
\centering
\vspace{-8pt}
\renewcommand{\arraystretch}{0.8}
\setlength{\tabcolsep}{5pt}
\setlength{\extrarowheight}{.0ex}
\resizebox{\textwidth}{!}{
\begin{tabular}{rr|r?r|r|r|r?r|r|r|r?r|r|r|r}
\toprule
& & & \multicolumn{4}{c?}{\textbf{Compressed size}} & \multicolumn{4}{c?}{\textbf{Compression time}} & \multicolumn{4}{c}{\textbf{Decompression time}} \\
& & \multicolumn{1}{c?}{\textbf{File}} & \multicolumn{1}{c|}{\tool{LZW}} & \multicolumn{1}{c|}{\tool{repair}} & \multicolumn{1}{c|}{\tool{zstd}} & \multicolumn{1}{c?}{\tool{lz4}} & \multicolumn{1}{c|}{\tool{LZW}} & \multicolumn{1}{c|}{\tool{repair}} & \multicolumn{1}{c|}{\tool{zstd}} & \multicolumn{1}{c?}{\tool{lz4}} & \multicolumn{1}{c|}{\tool{LZW}} & \multicolumn{1}{c|}{\tool{repair}} & \multicolumn{1}{c|}{\tool{zstd}} & \multicolumn{1}{c}{\tool{lz4}} \\
\midrule

\parbox[t]{2mm}{\multirow{6}{*}{\rotatebox[origin=c]{90}{\resizebox{78pt}{!}{\textbf{Uncompressed size}}}}} & \parbox[t]{2mm}{\multirow{3}{*}{\rotatebox[origin=c]{90}{\scriptsize\textbf{1 MB}}}} & \textit{Logs} & \loser{0.19} & \second{0.08} & \winner{0.07} & 0.12 & \second{0.04} & 0.19 & \loser{0.51} & \winner{0.03} & \loser{0.02} & 0.01 & \second{0.01} & \winner{0.004}  \\
& & \textit{Subtitles} & \loser{0.36} & \second{0.13} & \winner{0.11} & 0.15 & \second{0.04} & 0.25 & \loser{0.3} & \winner{0.03} & \loser{0.02} & 0.01 & \second{0.01} & \winner{0.004}  \\
& & \textit{Books} & 0.42 & \second{0.34} & \winner{0.27} & \loser{0.43} & \winner{0.04} & 0.29 & \loser{0.42} & \second{0.08} & \loser{0.02} & 0.02 & \second{0.01} & \winner{0.004}  \\
\cmidrule{2-15}
 & \parbox[t]{2mm}{\multirow{3}{*}{\rotatebox[origin=c]{90}{\scriptsize\textbf{500 MB}}}} & \textit{Logs} & \loser{96} & \second{38} & \winner{33} & 65 & \second{16.9} & 123.2 & \loser{819.1} & \winner{13.3} & \loser{7.8} & 5.5 & \second{1.1} & \winner{0.64}  \\
& & \textit{Subtitles} & \loser{191} & \second{66} & \winner{55} & 114 & \winner{19.9} & 169.3 & \loser{415.2} & \second{22.8} & \loser{8.6} & 8.2 & \second{1.2} & \winner{0.81}  \\
& & \textit{Books} &206 & \second{153} & \winner{129} & \loser{216} & \winner{20.2} & 198.6 & \loser{646.3} & \second{40.6} & 8.6 & \loser{9.7} & \second{2.0} & \winner{0.8}  \\
\bottomrule
\end{tabular}}
\vspace{5pt}
\caption{Sizes (in MB) of the compressed files and (de)compression times (in seconds). Maximum compression levels enabled.
(Blue = best; bold black = second best; red = worst).}
\label{table:compression}
\vspace{-10pt}
\end{table}

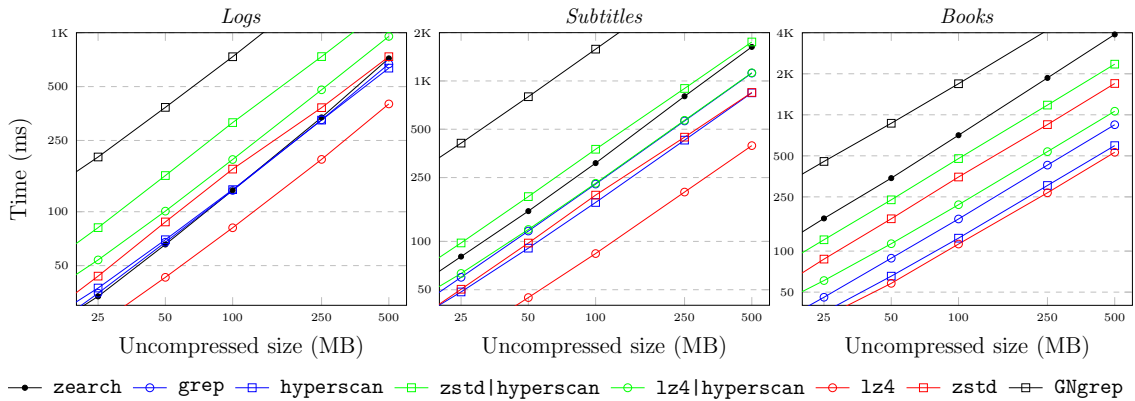
\begin{figure}[!ht]
\vspace{-10pt}
\resizebox{\textwidth}{!}{
\begin{tikzpicture}
  \begin{groupplot}[
      group style={
          group name=my plots,
          group size=3 by 1,
          xlabels at=edge bottom,
          ylabels at=edge left,
          horizontal sep=0.6cm,
          vertical sep=1cm,
          },
      xtick={1,5,10,25,50,100,250,500},
      ymajorgrids=true,
      xlabel near ticks,
      ylabel near ticks,
      xlabel={Uncompressed size (MB)},
      ylabel={Time (ms)},
      grid style=dashed,
      width=0.5\textwidth,
      xmode=log,
      ymode=log,
      log ticks with fixed point,
      tick label style={font=\tiny},
      legend columns=-1,
      legend style={draw=none,column sep=0.7ex},
  ]
\nextgroupplot[xmin=20, xmax=600, ymax=1000, ymin=30, yticklabels={15,25,50,100,250,500,1K,2K,4K,8K}, ytick={15,25,50,100,250,500,1000,2000,4000,8000}, legend to name=testLegend, title={\small \emph{Logs}}, title style={yshift=-7pt}]
\addplot[color=black, mark=*, mark size=1.3pt] coordinates{(1,1)};\addlegendentry{{\small\tool{zearch}}};
\addplot[color=blue, mark=o] coordinates{(1,1)};\addlegendentry{{\small\tool{grep}}};
\addplot[color=blue, mark=square] coordinates{(1,1)};\addlegendentry{{\small\tool{hyperscan}}};
\addplot[color=green, mark=square] coordinates{(1,1)};\addlegendentry{{\small\tool{zstd|hyperscan}}};
\addplot[color=green, mark=o] coordinates{(1,1)};\addlegendentry{{\small\tool{lz4|hyperscan}}};
\addplot[color=red, mark=o] coordinates{(1,1)};\addlegendentry{{\small\tool{lz4}}};
\addplot[color=red, mark=square] coordinates{(1,1)};\addlegendentry{{\small\tool{zstd}}};
\addplot[color=black, mark=square] coordinates{(1,1)};\addlegendentry{{\small\tool{GNgrep}}};
\addplot[color=black, mark=*, mark size=1.3pt] coordinates{(1,4.162)(5,9.442)(10,16.238)(25,33.638)(50,65.592)(100,131.692)(250,336.371)(500,718.913)};
\addplot[color=blue, mark=o] coordinates{(1,3.621)(5,8.654)(10,15.333)(25,35.058)(50,67.537)(100,131.483)(250,326.021)(500,669.133)};
\addplot[color=blue, mark=square] coordinates{(1,6.05)(5,11.262)(10,18.108)(25,37.467)(50,69.746)(100,132.85)(250,324.925)(500,633.954)};
\addplot[color=green, mark=square] coordinates{(1,8.696)(5,20.567)(10,35.438)(25,81.492)(50,159.208)(100,314.675)(250,735.458)(500,1430.696)};
\addplot[color=green, mark=o] coordinates{(1,7.083)(5,15.754)(10,25.133)(25,53.775)(50,100.746)(100,195.75)(250,479.75)(500,953.683)};
\addplot[color=red, mark=o] coordinates{(1,3.667)(5,7.667)(10,11.667)(25,23.333)(50,43.0)(100,81.333)(250,196.0)(500,400.0)};
\addplot[color=red, mark=square] coordinates{(1,4.667)(5,10.0)(10,18.333)(25,43.667)(50,87.667)(100,173.0)(250,381.333)(500,734.667)};
\addplot[color=black, mark=square] coordinates{(1,27.908)(5,57.292)(10,93.279)(25,202.146)(50,382.688)(100,734.075)(250,1810.258)(500,3802.442)};

\nextgroupplot[xmin=20, xmax=600, ymax=2000, ymin=40, yticklabels={15,25,50,100,250,500,1K,2K,4K,8K}, ytick={15,25,50,100,250,500,1000,2000,4000,8000}, title={\small \emph{Subtitles}}, title style={yshift=-5pt}]
\addplot[color=black, mark=*, mark size=1.3pt] coordinates{(1,5.608)(5,17.029)(10,33.975)(25,80.425)(50,154.429)(100,307.608)(250,802.433)(500,1629.858)};
\addplot[color=blue, mark=o] coordinates{(1,4.917)(5,13.783)(10,24.942)(25,59.725)(50,116.021)(100,227.662)(250,563.188)(500,1119.308)};
\addplot[color=blue, mark=square] coordinates{(1,6.592)(5,13.571)(10,22.408)(25,48.421)(50,90.763)(100,174.846)(250,427.65)(500,843.35)};
\addplot[color=green, mark=square] coordinates{(1,9.213)(5,23.221)(10,41.746)(25,97.817)(50,190.213)(100,375.288)(250,897.696)(500,1749.292)};
\addplot[color=green, mark=o] coordinates{(1,7.754)(5,17.833)(10,29.475)(25,63.15)(50,118.667)(100,230.312)(250,568.638)(500,1124.304)};
\addplot[color=red, mark=o] coordinates{(1,3.667)(5,8.0)(10,12.667)(25,24.0)(50,44.667)(100,84.0)(250,203.0)(500,395.333)};
\addplot[color=red, mark=square] coordinates{(1,4.333)(5,10.667)(10,21.0)(25,50.333)(50,97.667)(100,194.333)(250,446.333)(500,847.333)};
\addplot[color=black, mark=square] coordinates{(1,35.446)(5,96.308)(10,175.008)(25,409.642)(50,797.646)(100,1578.1)(250,3895.021)(500,7771.454)};

\nextgroupplot[xmin=20, xmax=600, ymax=4000, ymin=40, yticklabels={15,25,50,100,250,500,1K,2K,4K,8K}, ytick={15,25,50,100,250,500,1000,2000,4000,8000}, title={\small \emph{Books}}, title style={yshift=-5pt}]
\addplot[color=black, mark=*, mark size=1.3pt] coordinates{(1,9.688)(5,34.879)(10,67.888)(25,173.675)(50,342.996)(100,708.104)(250,1864.617)(500,3887.012)};
\addplot[color=blue, mark=o] coordinates{(1,3.967)(5,10.675)(10,19.717)(25,45.85)(50,88.692)(100,172.05)(250,427.342)(500,845.658)};
\addplot[color=blue, mark=square] coordinates{(1,5.796)(5,10.838)(10,17.504)(25,35.638)(50,65.487)(100,124.412)(250,302.933)(500,593.975)};
\addplot[color=green, mark=square] coordinates{(1,9.8)(5,26.608)(10,49.242)(25,121.112)(50,237.808)(100,477.337)(250,1178.146)(500,2351.9)};
\addplot[color=green, mark=o] coordinates{(1,6.875)(5,17.425)(10,28.337)(25,60.808)(50,113.312)(100,218.917)(250,536.704)(500,1061.717)};
\addplot[color=red, mark=o] coordinates{(1,4.0)(5,10.333)(10,15.667)(25,32.667)(50,58.0)(100,112.667)(250,267.667)(500,530.667)};
\addplot[color=red, mark=square] coordinates{(1,6.0)(5,17.0)(10,33.667)(25,87.333)(50,172.667)(100,349.667)(250,847.333)(500,1697.0)};
\addplot[color=black, mark=square] coordinates{(1,37.821)(5,103.808)(10,189.812)(25,454.75)(50,866.688)(100,1690.279)(250,4179.625)(500,8372.242)};

\end{groupplot}
\end{tikzpicture}
}
\begin{nscenter}
\resizebox{\textwidth}{!}{
\pgfplotslegendfromname{testLegend} 
}
\end{nscenter}

\caption{Average running time required to report the number of lines matching a regular expressions in a file and time required for decompression.
Colors indicate whether the tool performs the search on the uncompressed text (blue); the compressed text (black); the output of the decompressor (green); or decompresses the file without searching (red).
}\label{fig:comparison}
\vspace{-10pt}
\end{figure}

We first run each experiment 3 times as warm up so that the files are loaded in memory.
Then we measure the running time 30 times and compute the \emph{confidence interval} (with 95\% confidence) for the running time required to count the number of matching lines for a regular expression in a certain file using a certain tool.
We consider the \emph{point estimate} of the confidence interval and omit the \emph{margin of error} which never exceeds the $9\pct$ of the point estimate for the reported experiments.
Figure~\ref{fig:comparison} summarizes the obtained results when considering, for all files, the regular expressions: ``\verb!what!'', ``\verb!HTTP!'', ``\verb!.!'', ``\verb!I .* you !'', ``\verb! [a-z]{4} !'', ``\verb! [a-z]*[a-z]{3} !'', ``\verb![0-9]{4}!'', ``\verb![0-9]{2}/(Jun|Jul|Aug)/[0-9]{4}!''.
For clarity, we report only on the most relevant tools among the ones considered.
For \tool{lz4} and \tool{zstd}, we report the time required to decompress the file and send the output to \tool{/dev/null}.

\SubSection{Analysis of the Results.}
Figure~\ref{fig:comparison} and Table~\ref{table:compression} show that the performance of \tool{zearch} improves with the compression ratio.
This is to be expected since \tool{zearch} processes each grammar rule exactly once and better compression results in less rules to be processed.
In consequence, \tool{zearch} is the fastest tool for counting matching lines in compressed \emph{Log} files while it is the second slowest one for the \emph{Books}.

In particular, \tool{zearch} is more than $25\pct$ faster than any other tool working on compressed \emph{Log} files.
Actually \tool{zearch} is competitive with \tool{grep} and \tool{hyperscan}, even though these tools operate on the uncompressed text.
These results are remarkable since \tool{hyperscan}, unlike \tool{zearch}, uses algorithms specifically designed to take advantage of SIMD parallelization.\footnote{According to the documentation, \tool{hyperscan} \emph{requires}, at least, support for SSSE3.}

Finally, the fastest tool for counting matching lines in compressed \emph{Subtitles} and \emph{Books}, \tool{lz4|hyperscan}, applies to files larger than the ones obtained when compressing the data with \tool{repair} (see Table~\ref{table:compression}).
However, when considering a better compressor such as \tool{zstd}, which achieves slightly more compression than \tool{repair}, the decompression becomes slower.
As a result, \tool{zearch} outperforms \tool{zstd|hyperscan} by more than $7\pct$ for \emph{Subtitles} files and $50\pct$ for \emph{Logs}.

\Section{Fine-Grained Complexity}\label{sec:complexity}

The grammars produced by \tool{repair} break the definition of SLP in behalf of the compression by allowing the axiom rule to have more than two symbols on the right hand side.
This is due to the fact that the axiom rule is built with the remains of the input text after creating all grammar rules.
Typically $\len{σ} \geq \len{\pr}$ so the way in which the axiom is processed heavily influences the performance of \tool{zearch}.

Furthermore, our experiments show that the performance of \tool{zearch} is typically far from its worst case complexity.
This is because the worst case scenario assumes each string generated by a grammar variable labels a path between each pair of states of the automaton.
However, we only observed such behavior in contrived examples.

\SubSection{Processing the Axiom Rule.}

Algorithm~\ref{alg:algorithmTheoryCount} could process the axiom rule $X_{\len{V}}{→}σ$ by building an SLP with rules $\{S_{1}{→}(σ)_1(σ)_2\} \cup \{S_i{→}S_{i{-}1}(σ)_{i{+}1} \mid i = 2\ldots \len{σ}{-}2\} \cup \{X_{\len{V}}{→}S_{\len{σ}{-}2}(σ)_\dag\}$.
However it is more efficient to compute the set of states reachable from the initial ones when reading the string generated with $S_1$ and update this set for each symbol $(σ)_i$.
To perform the counting note that $\counting{S_i}$ is only used to compute $\counting{S_{i+1}}$ and can be discarded afterwards.
This yields an algorithm running in $\mathcal{O}\left(p{\cdot} s^3{+}\len{σ}{\cdot} s^2\right)$ time using $\mathcal{O}\left(p{\cdot} s^2\right)$ space where $p$ is the number of rules of the input grammar and $X_{\len{V}}{→}σ$ its axiom.

\SubSection{Complexity in Terms of Operations Performed by the Algorithm}
Define $s_{τ,q}=\len{\{q' \mid (q,τ,q') \in δ\}}$ and $s_τ = \sum_{q \in Q}s_{τ,q}$.
Let us recall the complexity of Algorithm~\ref{alg:algorithmTheoryCount} according to the described data structures.
The algorithm iterates over the $p$ rules of the grammar and, for each of them, \begin{inparaenum}
\item [1)] initializes matrix $\mathcal{N}$ with $s_{β_\ell}$ elements\footnote{We need to set up to $s$ sentinel values for the rows in $\mathcal{N}$ not used for storing $s_{β_\ell}$} and
\item [2)] iterates through $\mathcal{N}[q'][0…s_{β_\ell,q'}]$ for each pair $(q_1,q') \in \edges{α_\ell}$.  
\end{inparaenum}
Then it processes the axiom rule iterating, for each symbol $(σ_i)$, through $s_{(σ)_i}$ transitions.
These are all the operations performed by the algorithm with running time dependent on the size of the input.
Hence, Algorithm~\ref{alg:algorithmTheoryCount} runs in $\mathcal{O}\left(\sum_{\ell=1}^{\len{V}}\tilde{s}_\ell+\sum_{i=1}^\len{σ}s_{(σ)_i} \right)$ time where $\tilde{s}_\ell=s_{β_\ell}+s+\sum_{(q_1,q') \in \edges{α_\ell}}\left(1{+}s_{β_\ell,q'}\right)$.
Note that $\tilde{s}_\ell \leq s^3$ and $s_{(σ)_i} \leq s^2$.

\begin{table}[!ht]
\centering
\resizebox{\textwidth}{!}{
\renewcommand{\arraystretch}{0.60}
\setlength{\tabcolsep}{4pt}
\setlength{\extrarowheight}{1ex}
\small
\begin{tabular}{lr?r|rrrrr?r|rrrrr}
\toprule
\multicolumn{1}{c}{\multirow{2}{*}{\textbf{Expression}}} & \multicolumn{1}{c?}{\multirow{2}{*}{$s$}}  & \multicolumn{1}{c|}{\multirow{2}{*}{$s^3$}} & \multicolumn{5}{c?}{percentiles for $\tilde{s}_{\ell}$} & \multicolumn{1}{c|}{\multirow{2}{*}{$s^2$}} & \multicolumn{5}{c}{percentiles for $s_{(σ)_i}$} \\
\multicolumn{1}{c}{} & \multicolumn{1}{c?}{} & \multicolumn{1}{c|}{} & {\footnotesize $50$\pct} & {\footnotesize $75$\pct} & {\footnotesize $95$\pct} & {\footnotesize $98$\pct} & {\footnotesize$100$\pct} & \multicolumn{1}{c|}{} & {\footnotesize $50$\pct} & {\footnotesize $75$\pct} & {\footnotesize $95$\pct} & {\footnotesize $98$\pct} & {\footnotesize$100$\pct}\\
\midrule
{\small ``\tool{I .* you}''} & 8 & 512 & 3 & 13 & 15 & 17 & 28 & 64 & 3 & 3 & 5 & 5 & 8 \\
{\small ``\tool{.*[A-Za-z ]\{5\}}''} & 7 & 343 & 14 & 25 & 48 & 48 & 48 & 49 & 11 & 14 & 14 & 14 & 14 \\
{\small ``\tool{.*[A-Za-z ]\{10\}}''} & 12 & 1728 & 29 & 51 & 86 & 95 & 98 & 144 & 16 & 26 & 29 & 29 & 29 \\
{\small ``\tool{.*[A-Za-z ]\{20\}}''} & 22 & 10648 & 57 & 87 & 132 & 153 & 198 & 484 & 23 & 38 & 52 & 58 & 59 \\
{\small ``\tool{(((((.)*.)*.)*.)*.)*}''} & 6 & 216 & 12 & 29 & 209 & 209 & 209 & 36 & 29 & 29 & 29 & 29 & 29 \\
\bottomrule
\end{tabular}
}
\vspace{5pt}
\caption{Analysis of the values $\tilde{s}_\ell$ and $s_{(σ)_i}$ obtained when considering different regular expressions to search \emph{Subtitles} (100 MB uncompressed long).
The fifth column of the first row indicates that when considering the expression ``\tool{I .* you}'', for 75\% of the grammar rules we have $\tilde{s}_\ell \leq 5$ while $s^3=512$.}
\label{table:Behavior}
\vspace{-10pt}
\end{table}

In the experiments we observed that $\tilde{s}_\ell$ and $s_{(σ_i)}$ are usually much smaller than $s^3$ and $s^2$, respectively, as reported in Table~\ref{table:Behavior}.
Indeed, \tool{zearch} exhibits almost linear behavior with respect to the size of the FSA built from the expression.
Nevertheless, there are regular expressions that trigger the worst case behavior (last row in Table~\ref{table:Behavior}), which cannot be avoided due to the result of Abboud et al.~\cite{Amir2018FineGrained} describe before.

\Section{Related Work}%
\label{sec:related}
Regular expression matching on compressed text is a well-studied problem consisting on deciding whether a string generated by an SLP matches a regular expression (represented as an automaton).
Plandowsky et al.~\cite{plandowski1999complexity} reduced the problem to a series of matrix multiplications, showing it can be solved in $\mathcal{O}(ps^3)$ time ($\mathcal{O}(ps)$ for deterministic automata) where $p$ is the size of the grammar and $s$ is the size of the automaton.
Independently, Esparza et al.~\cite{esparza2000uniform} defined an algorithm to solve a number of decision problems involving automata and context-free grammars which, when restricted to SLPs, results in a particular implementation of Plandowsky's approach.
Removing the function \textsc{count} from Algorithm~\ref{alg:algorithmTheoryCount} and replacing the \emph{return} statement of the algorithm by $\left((I\times \{X_\len{V}\}\times F) \cap δ: \True\ \mathbin{?} \ \False\right)$ we are left with an efficient implementation of Plandowsky's approach.

The first algorithm searching grammar-compressed text for regular expression matches is due to Navarro~\cite{navarro2003regular}, \tool{GNgrep}, and it is defined for LZ78/LZW compressed text.
His algorithm reports all $k$ positions in the uncompressed text at which a substring matching the expression ends (but not the matches) in $\mathcal{O}(2^s{+}s{\cdot} p{+}k{\cdot} s{\cdot} \log{s})$ time using $\mathcal{O}(2^s{+}p{\cdot} s)$ space.
The algorithm computes the number of matches by enumerating these positions.
To the best of our knowledge this is the only algorithm searching compressed text for regular expression matches that has been implemented and evaluated in practice.
Bille et al.~\cite{bille2009improved} improved the result of Navarro by defining a data structure of size $o(p)$ to represent LZ78 compressed texts that offers a time-space trade off (reflected in parameter $τ$) for finding all occurrences of a regular expression in a LZ78 compressed text.
Their algorithm operates in $\mathcal{O}(p{\cdot} s{\cdot} (s{+}\tau){+}k{\cdot} s{\cdot} \log{s})$ time using $\mathcal{O}(p{\cdot} s^2/\tau {+} p{\cdot} s)$ space, where $1 \leq τ \leq p$. 
Note that these approaches to search compressed text exhibit running time linear in the size of the uncompressed text which might be exponentially larger than its compressed version.
In consequence, they are not optimal nor competitive with the state of the art, as show by Figure~\ref{table:Behavior}.

\Section{Conclusions and Future Work}\label{sec:conclusions}
We have presented the first algorithm for \emph{counting} the number of lines in a grammar-compressed text containing a match for a regular expression.
The algorithm applies to any grammar-based compression scheme and is nearly optimal for regular expression matching on compressed text.
Furthermore, we described the data structures required to achieve nearly optimal complexity and used them to implement a (sequential) tool that significantly outperforms the (parallel) state of the art to solve this problem.
Indeed, when the grammar-based compressor achieves hight compression ratio, which is the case for automatically generated \emph{Log} files, \tool{zearch} uses up to $25\pct$ less time than \tool{lz4|hyperscan}, even outperforming \tool{grep} and being competitive with \tool{hyperscan}.

Finally, Algorithm~\ref{alg:algorithmTheoryCount} allows for a conceptually simple parallelization since any set of rules such that no variable appearing on the left hand side of a rule appears on the right hand side of another, can be processed simultaneously.
Indeed, a theoretical result by Ullman et al.~\cite{ullman1988parallel} on the parallelization of Datalog queries can be used to show that counting the number of lines in a grammar-compressed text containing a match for a regular expression is in $\mathcal{NC}^2$ when the automaton built from the expression is acyclic.
We are optimistic about the possibilities of a parallel version of \tool{zearch}.

\Section{References}
\bibliographystyle{IEEEbib}

\end{document}

%% file: dataStructure.tex
\draw[shift={(285.368, 716.58)}, scale=0.6563]
    (0, 0) rectangle (32, -32);
  \draw[shift={(306.37, 716.58)}, scale=0.6563]
    (0, 0) rectangle (32, -32);
  \draw[shift={(285.368, 695.577)}, scale=0.6563]
    (0, 0) rectangle (64, -32);
  \node[ipe node, font=\large]
     at (188.695, 690.754) {$\counting{X_i}$};
  \draw[shift={(176, 711.737)}, xscale=0.8341, yscale=0.9892]
    (0, 0) rectangle (48, -32);
  \draw[shift={(123.816, 725.609)}, xscale=0.7558, yscale=0.7487]
    (0, 0) rectangle (16, -16);
  \draw[shift={(123.816, 701.651)}, xscale=0.7558, yscale=0.7487]
    (0, 0) rectangle (16, -16);
  \draw[shift={(123.816, 677.693)}, xscale=0.7558, yscale=0.7487]
    (0, 0) rectangle (16, -16);
  \node[ipe node]
     at (113.401, 716.068) {$0$};
  \node[ipe node]
     at (115.793, 692.139) {$i$};
  \node[ipe node]
     at (89.623, 667.626) {$p{+}\len{Σ}$};
  \draw[ipe dash dashed]
    (131.5102, 702.6101)
     -- (175.901, 711.756);
  \draw[ipe dash dashed]
    (131.7285, 688.1998)
     -- (175.901, 679.767);
  \node[ipe node, font=\large]
     at (227.757, 690.888) {$\edges{X_i}$};
  \node[ipe node, font=\large]
     at (296.463, 683.341) {ptr};
  \node[ipe node, font=\large]
     at (289.156, 705.124) {$q_{i_1}$};
  \node[ipe node, font=\large]
     at (310.158, 705.124) {$q_{i_2}$};
  \node[ipe node, font=\large]
     at (479.878, 690.956) {NULL};
  \pic[ipe mark tiny]
     at (129.3755, 711.3358) {ipe disk};
  \pic[ipe mark tiny]
     at (129.3876, 707.4932) {ipe disk};
  \pic[ipe mark tiny]
     at (129.4054, 703.6406) {ipe disk};
  \pic[ipe mark tiny]
     at (129.3667, 687.1024) {ipe disk};
  \pic[ipe mark tiny]
     at (129.3876, 683.5351) {ipe disk};
  \pic[ipe mark tiny]
     at (129.3684, 679.9975) {ipe disk};
  \node[ipe node]
     at (167.443, 658.533) {$\mathcal{N}[i_1][0…2] = [i_2,i_4,{-}1]$};
  \draw[shift={(349.367, 716.58)}, scale=0.6563]
    (0, 0) rectangle (32, -32);
  \draw[shift={(370.37, 716.58)}, scale=0.6563]
    (0, 0) rectangle (32, -32);
  \draw[shift={(349.367, 695.577)}, scale=0.6563]
    (0, 0) rectangle (64, -32);
  \draw[shift={(327.381, 685.411)}, xscale=0.4477, yscale=0.4841, ->]
    (0, 0)
     -- (29.958, 0.037)
     -- (29.958, 20.86)
     -- (48.343, 20.954);
  \node[ipe node, font=\large]
     at (360.735, 683.116) {ptr};
  \node[ipe node, font=\large]
     at (354.359, 704.974) {$q_{i_3}$};
  \node[ipe node, font=\large]
     at (375.362, 704.974) {$q_{i_4}$};
  \draw[shift={(413.367, 716.58)}, scale=0.6563]
    (0, 0) rectangle (32, -32);
  \draw[shift={(434.37, 716.58)}, scale=0.6563]
    (0, 0) rectangle (32, -32);
  \draw[shift={(413.367, 695.577)}, scale=0.6563]
    (0, 0) rectangle (64, -32);
  \node[ipe node, font=\large]
     at (424.735, 683.116) {ptr};
  \node[ipe node, font=\large]
     at (418.359, 704.974) {$q_{i_1}$};
  \node[ipe node, font=\large]
     at (439.362, 704.974) {$q_{i_4}$};
  \draw[shift={(391.504, 685.815)}, xscale=0.4429, yscale=0.4629, ->]
    (0, 0)
     -- (29.958, 0.037)
     -- (29.958, 20.86)
     -- (48.343, 20.954);
  \draw[shift={(455.566, 685.358)}, xscale=0.46, yscale=0.4804, ->]
    (0, 0)
     -- (29.958, 0.037)
     -- (29.958, 20.86)
     -- (48.343, 20.954);
  \node[ipe node]
     at (317.726, 658.465) {$\mathcal{N}[i_3][0…1] = [i_4,{-}1]$};
  \node[ipe node]
     at (204.039, 725.877) {$\mathcal{M}[i_1][i_2] {=}\mathcal{M}[i_3][i_4] {=}
\mathcal{M}[i_1][i_4] {=} X_i$};
  \node[ipe node, font=\large]
     at (124.061, 727.992) {$\mathcal{A}$};
  \draw[->]
    (256.2657, 694.9724)
     -- (283.9817, 695.0154);
  \draw[shift={(216.062, 711.739)}, xscale=0.8341, yscale=0.9892]
    (0, 0) rectangle (48, -32);